\documentclass[sigconf]{acmart}

\usepackage{balance}
\usepackage{multirow}

\def\BibTeX{{\rm B\kern-.05em{\sc i\kern-.025em b}\kern-.08emT\kern-.1667em\lower.7ex\hbox{E}\kern-.125emX}}
    
\copyrightyear{2020}
\acmYear{2020}
\setcopyright{acmcopyright}\acmConference[ICMR '20]{Proceedings of the 2020 International Conference on Multimedia Retrieval}{June 8--11, 2020}{Dublin, Ireland}
\acmBooktitle{Proceedings of the 2020 International Conference on Multimedia Retrieval (ICMR '20), June 8--11, 2020, Dublin, Ireland}
\acmPrice{15.00}
\acmDOI{10.1145/3372278.3390691}
\acmISBN{978-1-4503-7087-5/20/06}

\begin{document}

\fancyhead{}

\title{Search Result Clustering in Collaborative Sound Collections}

\author{Xavier Favory, Frederic Font, Xavier Serra}
\email{name.surname@upf.edu}
\affiliation{%
  \institution{Music Technology Group \\
  Universitat Pompeu Fabra}
  \city{Barcelona}
  \country{Spain}
}

\begin{abstract}
The large size of nowadays' online multimedia databases makes retrieving their content a difficult and time-consuming task.
Users of online sound collections typically submit search queries that express a broad intent, often making the system return large and unmanageable result sets.
Search Result Clustering is a technique that organises search-result content into coherent groups, which allows users to identify useful subsets in their results.
Obtaining coherent and distinctive clusters that can be explored with a suitable interface is crucial for making this technique a useful complement of traditional search engines.
In our work, we propose a graph-based approach using audio features for clustering diverse sound collections obtained when querying large online databases.
We propose an approach to assess the performance of different features at scale, by taking advantage of the metadata associated with each sound.
This analysis is complemented with an evaluation using ground-truth labels from manually annotated datasets.
We show that using a confidence measure for discarding inconsistent clusters improves the quality of the partitions.
After identifying the most appropriate features for clustering, we conduct an experiment with users performing a sound design task, in order to evaluate our approach and its user interface.
A qualitative analysis is carried out including usability questionnaires and semi-structured interviews. 
This provides us with valuable new insights regarding the features that promote efficient interaction with the clusters.

\end{abstract}




\keywords{search interfaces; sound retrieval; sound clustering; unsupervised classification; audio features; neural-network embeddings \vspace{1em}}


\maketitle

\section{Introduction}

Sounds used in movies, video-games, music and other media often originate from sound collections.
Thanks to online sharing platforms, a wide variety of content can be easily and freely shared.
In these platforms, the collection is generated by its users instead of coming from professional studios.
This results in non-uniformly-annotated content compared to professional libraries, which involve experts for annotating and organizing the collections~\cite{font2018sound}.
Therefore, browsing collaborative online sound collections is more arduous.

Typically, users interact with these collections by using text-based search engines.
The primary role of these retrieval systems is to support users in accessing relevant content corresponding to their needs.
%
%
After entering a text query, a user often faces a long list of results.
In the absence of specific query terms, the system may be unable to differentiate the relevance of the retrieved sounds to the user, whose needs are often very precise and related to highly-specialised tasks.
The user might be looking for audio clips with distinctive and detailed characteristics that can be represented by a wide range of proprieties.
In sound design, for instance, a user could be searching for a door closing sound which has a grinding noise that fits well with the movie ambiance and the visual aspect of the door.
In the case of music creation, they may be interested in finding instrument loops playing in a certain tonality, at a specific tempo and with different timbres.
In order to locate sounds of interest, the user usually needs to check the results one by one, listening to some of them and judging their relevance.
The process of finding the most appropriate sounds for specific user needs can be very time consuming and can fail in retrieving some sounds of interest when interacting with huge collections.

Users can narrow down the search by reformulating their queries, or by using some facet-based and tag-based filters~\cite{tunkelang2009faceted}.
When available, a user might find it very informative to explore the text and tags accompanying the sounds in the results page, in order to identify new terms for reformulating the query.
Tag clouds that organise and display popular tags from the results are particularly useful for that purpose. The user can quickly find particular sub-topics present in the search results~\cite{sinclair2008folksonomy}.
Nonetheless, functionalities based on textual metadata depend critically on the quality of the annotations, which is often limited in collaborative collections.
For this reason, content-oriented methods that are based on the audio content itself have increased potential in the development of novel approaches to navigate search results in such collections.

To that end, one complementary feature that search engines can incorporate is Search Result Clustering, which consists of grouping the results into different labeled clusters or categories.
It enables the user to enter a weakly-specified query and then explore the different themes that have been automatically extracted from the query results.
Clustering engines can complement the search by providing a faster way to retrieve relevant items, facilitating topic exploration and preventing the overlook of information~\cite{carpineto2009survey}.
However, such systems depend on more than just the clustering algorithm.
In order to guide the user to locate relevant items in the different clusters, meaningful labels should be assigned to each of them.
Moreover, the clustering has to be performed online within a short response time, requiring high computational efficiency.
Finally, the clustering engine requires a graphical user interface that provides an intuitive way to navigate the clusters, e.g. by conveying visual information.

In this work, we present the development of an audio-based Search Result Clustering engine that integrates with the Freesound website~\cite{font2013freesound}. 
In Section 2, we provide an overview of the related work from the literature. 
We then introduce our graph-based clustering approach and our interface in Section 3.
In section 4, we compare the performance of different features taken from the literature, by using sound metadata and ground-truth labels from manually-annotated datasets.
Section 5 exposes an evaluation of the system with users performing a sound design task, which allows us to identify the key features that would enable the user to efficiently interact with the obtained clusters.
We end the paper with a conclusions section and discussion about future work.

\section{Related work}


In order to be able to organize and retrieve a large amount of poorly labeled data, automatic annotation methods have been extensively addressed in the research community.
The main requirement of these content-based approaches is a reliable numerical feature that can represent the content. 
These features were originally developed by carefully designing low-level descriptors by relying on domain knowledge about the sound class invariances.
Some are derived directly from the time domain audio representation, whereas others use a spectral representations of the sounds which is mostly motivated by the fact that human perception widely relies on the frequency content of sound signals. 
Combining these different types of features enables the representation of the timbre of musical instruments or more high-level characteristics such as music moods and genres~\cite{peeters2011timbre, herrera2003automatic, kim2010music, eronen2000musical, tzanetakis2002musical}.

Recently, techniques using Artificial Neural Networks have been able to provide an alternative to the handcrafted features previously developed. 
Tasks, such as auto-tagging or classification can be performed directly from the raw audio~\cite{pons2017end}, or using a spectrogram representation~\cite{hershey2017cnn}.
The internal representation that neural networks learn on one task, can be used for other applications~\cite{choi2017transfer}.
These transfer learning approaches make use of pre-trained models as a starting point for different tasks.
First layers from trained neural networks often learn similar features which can be applicable for many datasets and tasks~\cite{yosinski2014transferable}.
Intermediate layers can serve as a higher-level representations which can be used for clustering~\cite{jansen2017large}.


%

Clustering is a type of unsupervised classification which consists in organising similar objects in groups called clusters.
Regardless of the clustering method, the content similarity measure involved is fundamental to the definition of a cluster.
This similarity notion is often derived from a feature space, on which a numerical distance or similarity is calculated.
When clustering audio content, the features and distance measures are often chosen carefully for a given specific task~\cite{black1997automatically, martins2007polyphonic, niessen2013hierarchical}.
However in the context of large online collections, the content is very diverse, containing speech, musical or environmental sounds.
This makes the choice of features and distance metric even more challenging.
Among the different approaches for clustering~\cite{fahad2014survey, xu2015comprehensive}, in the context of multimedia documents, density-based algorithms such as graph-based clustering methods are particularly well designed for dealing with computational efficiency and the heterogeneous aspect of the data~\cite{petkos2017graph}.
Moreover, in the context of sounds, graph-based algorithms based on k-Nearest Neighbors have been shown to be scalable and able to adapt to areas of different densities~\cite{roma2015algorithms}. 

Signal processing approaches and machine learning techniques have been employed for organising and visualising large amounts of audio content~\cite{neumayer2005content, fallgren2018tool, wongsuphasawat2015voyager, tzanetakis2001marsyas3d}.
These techniques often make use of a dimensionality reduction technique over numerical features, and project the content into a 2-dimensional space where similar sounds are close from each other.
The user can locate a sound of interest and then explore its neighborhood.
For instance, spaces conveying timbral characteristics of the sounds enable quick exploration of collections~\cite{tzanetakis2001marsyas3d}.
The use of visualisation tools in browsing systems have been shown to facilitate and encourage broader exploration~\cite{wongsuphasawat2015voyager}.
Clustering can be used in this space in order to automatically extract groups of similar sounds or music~\cite{neumayer2005content}.
However, to our knowledge, this is the first study describing the integration of a sound clustering algorithm in a search engine. 



\section{Proposed Approach}

\begin{figure}
 \centering
 \includegraphics[width=\columnwidth]{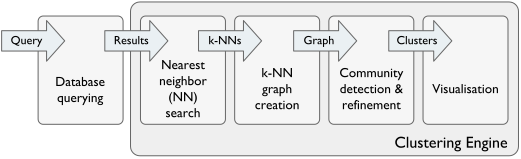}
 \caption{Diagram representing the steps of our clustering engine.}
 \label{fig:src_graph}
\end{figure}

Our clustering engine consists of several steps illustrated in Figure~\ref{fig:src_graph}. 
After collecting the results retrieved by the user's query, (i) nearest neighbor searches are computed for the top ranked results, (ii) a K-Nearest Neighbors graph is created, (iii) a community detection algorithm assigns each sound to one cluster with possible extra refinements, (iv) and finally the results are displayed with a visualisation.
The method is available as a service and its integration within the Freesound search engine is currently under development.
The code is available at this repository: \url{http://omitted.for.blind.review}.
We describe the audio features we will be using and comparing (Section 3.1), the graph-based clustering method (Section 3.2), a refinement strategy for discarding inconsistent clusters (Section 3.3) and the user interface of the system (Section 3.4).

\subsection{Audio Features}
In this work, we compare the performance of our clustering method using three different sets of features. One set of manually selected features motivated from the literature (F1)~\cite{peeters2011timbre, herrera2003automatic, kim2010music, eronen2000musical, tzanetakis2002musical}, another set contains all of the lowlevel features available from the Essentia Freesound Extractor (F2)~\cite{bogdanov2013essentia}, and the third one uses embeddings from a neural network model trained on AudioSet~\cite{hershey2017cnn, gemmeke2017audio} (F3).
Table \ref{tab:features} details the different features.

Most of the traditional acoustic features (F1, F2) are computed on frames of approximately 50 ms.
These frame-based features are summarised into a single feature vector, which ignores the temporal order.
It includes minimum and maximum values, mean and variance of the direct features and of their first and second derivatives.
The rest of the features, e.g. \textit{logattacktime}, consist of a single numerical value for an entire audio clip.
A dimentionality reduction is then performed using Principal Component Analysis over the entire sound collection, to reduce these concatenated statistics and values into a feature vector with 100 dimensions.
%
The deep neural network embeddings (F3) are calculated on windows of approximately 1 second.
These frame-based features are then aggregated with the mean statistic only.
%

\setlength{\tabcolsep}{0.3em}
\begin{table}
\small
 \begin{center}
 \begin{tabular}{c|l}
  \toprule
  Feature sets & Features \\
  \midrule
  \multirow{7}{*}{F1} & spectral centroid / complexity / spread / energy \\
  & energyband high / skewness /flatness db / rolloff, \\
  & temporal decrease / spread / kurtosis / skewness / \\
  & centroid, logattacktime, strongdecay, \\
  & effective duration, zerocrossingrate, \\
  & tristimulus, mfcc, dissonance \\
  \\
  \multirow{2}{*}{F2} & lowlevel features from the Essentia Freesound Extractor \\
  & \url{https://freesound.org/docs/api/analysis_docs.html} \\
  \\
  \multirow{2}{*}{F3} &  embeddings from AudioSet pre-trained model \\
  & \url{https://git.io/fjqsV} \\
  \bottomrule
 \end{tabular}
\end{center}
 \caption{The different features compared in this work.}
 \label{tab:features}
\end{table}

\subsection{Graph-Based Clustering}

Instead of directly using the features as input of a clustering method, we construct an intermediate representation of the data using a k-Nearest Neighbor Graph~\cite{dong2011efficient}. 
Each vertex represents a sound, and undirected edges connect each sound to its $k$ most similar according to the euclidean distance.
Some preliminary empirical tests made us choose $log2(N)$ for the value of $k$, where $N$ is the number of elements to cluster.
This allows us to reach a sufficient number of neighbors for small collections, while limiting it for larger collections, which ensures low-computational complexity.
Then, we use a community detection algorithm based on modularity optimisation for finding a partition of the graph~\cite{blondel2008fast}.
There are several reasons why we use such an approach. 
First it has been shown to be able to find clusters of different densities~\cite{roma2012factors}.
Second, the number of clusters does not need to be specified.
Also, it can take advantage of nearest neighbors search techniques that can be fast to compute (e.g. \cite{cayton2008fast} or similar approximate methods~\cite{aumuller2019ann, indyk1998approximate}).
In our work we use the Gaia library for performing nearest neighbor searches (\url{https://github.com/MTG/gaia}).
Another advantage of these graph-based methods is their simplicity which allows to use some interpretable heuristics for modifying the graph or its partition like discarding clusters of low quality.

\subsection{Discarding low quality clusters}

The amount of intra-cluster and inter-cluster edges (which are related to the modularity definition~\cite{newman2004finding}) can be used for defining an internal quality metric which is only based on data used by the clustering algorithm. 
Since we use the same data representation for the quality metric and for the clustering algorithm, it is not clear if it can be used for automatically assessing the quality of a cluster in terms of compactness and distinctiveness.
In this work, we are interested in investigating its use as a confidence score for quantising the quality of an individual cluster, and possibly discard low quality clusters that should not be presented to the user in the context of Search Result Clustering.
Our confidence score $c$ of a cluster ranges from 0 to 1 and is defined as following:
\begin{equation}
c = \frac {number \ of \ intra \ cluster \ edges} {total \ number \ of \ inter \ \& \ intra \ cluster \ edges}
\end{equation}
This confidence score will be higher for clusters that are more coherent, i.e., the sounds within a cluster are more similar to sounds within the same cluster than to sounds from other clusters.
In the case that many of the elements of a cluster have edges to elements of other clusters, this score will be lower.
This score penalises clusters that are not compact and distinct from other clusters.
In this work, we investigate the use of this simple internal metric (which does not make use of any external knowledge about the data) as a confidence measure for discarding potentially irrelevant clusters.

\subsection{User Interfaces}
\begin{figure*}
 \centering
 \includegraphics[width=0.93\textwidth]{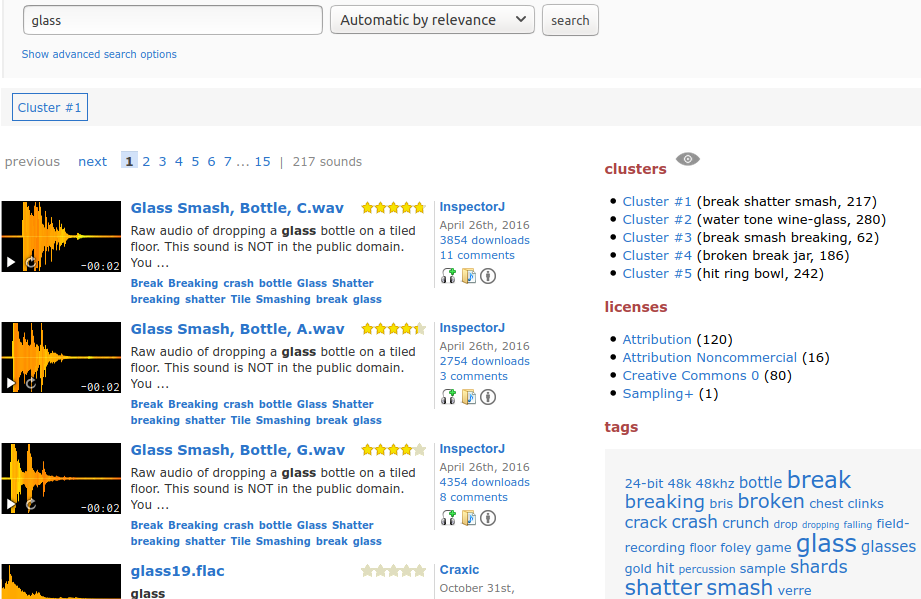}
 \caption{Page displaying the result of the query \textit{glass} of the \textit{cluster \#1}. Clicking on a cluster facet on the right applies a cluster filter. Three labels are shown for each cluster, together with the number of sounds they contain.}
 \label{fig:src_facet}
\end{figure*}

\begin{figure*}
 \centering
 \includegraphics[width=0.93\textwidth]{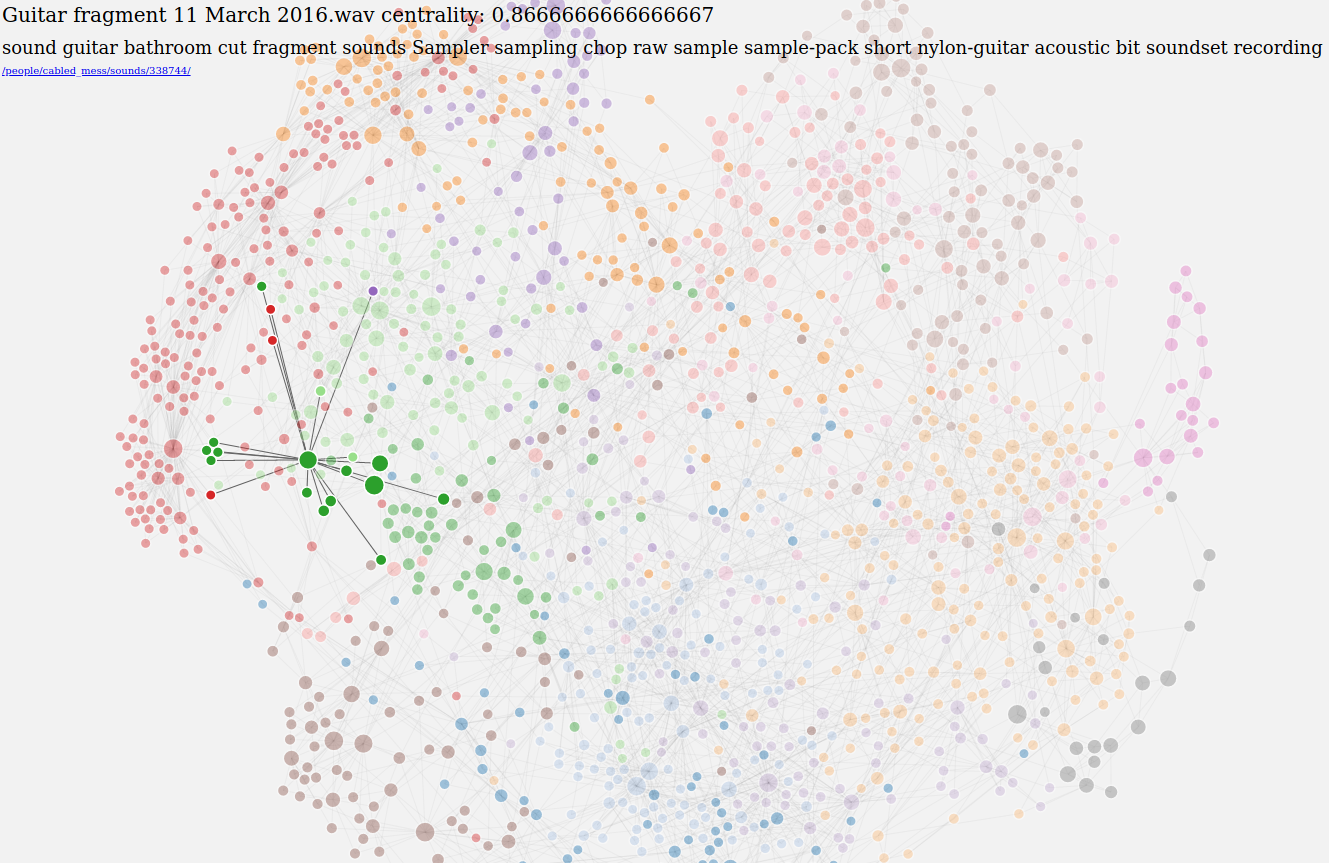}
 \caption{The graphical 2D visualisation of sounds retrieved with the query \textit{guitar}. Each circle represents a sound. Placing the mouse on one will play the associated sound. Clicking on it displays some information at the top of the screen and highlights neighbor nodes.}
 \label{fig:src_2d_visu}
\end{figure*}

To allow the user to interact with the clusters, we propose two different interfaces.
One consists of a traditional facet filtering approach, where the user can apply filters on the result to display only sounds from one cluster. Figure \ref{fig:src_facet} shows the modified Freesound search interface with the added clusters facets.
Three labels are displayed for each cluster which correspond to the most occurring tags in the cluster.
The second interface consists of a 2D-visualisation of the k-Nearest Neighbor Graph, where colors are used for representing clusters as shown in Figure \ref{fig:src_2d_visu}. Sounds can be played by hovering the mouse on the nodes.
Moreover, clicking on a node will highlight its neighbors in order to ease neighborhood exploration. 

\section{Feature Performance comparison}

In this section we show some comparative performance results of our approach using the different sets of features previously described.
We perform two different evaluations: one using internal validation and another using external validation.
For the first one, we propose to leverage information from an existing sound sharing website to automatically evaluate the clustering performances at scale.
We then perform a standard evaluation which uses ground truth labels taken from manually annotated datasets.
One of the goals to perform these two evaluations is to validate that our first evaluation that does not require known ground truth labels is adequate for comparing the performance of different clustering methods.

\subsection{Internal Validation}

We consider the Freesound database as a use case and we perform clustering on the search results of popular queries submitted by real users of the platform. We focus only on sounds with duration from 0 to 10 seconds.
For a quantitative evaluation, we make use of the sounds' metadata that is provided by the creator of the content in the Freesound platform.

Evaluating a clustering automatically is a complicated task, and there are different types of metrics that can be used~\cite{manning2010introduction}.
Some of them are referred as internal metrics, and they are used when no ground truth label is known~\cite{liu2010understanding}.
The Calinski-Harabasz Index (CHI)~\cite{calinski1974dendrite} evaluates the cluster validity based on the average between- and within-cluster sum of squares as shown in this equation:
\begin{equation}
CH(k)=[B(k) / W(k)] × [(n - k)/(k - 1)]
\end{equation}
Where $n$ corresponds to the number of data points, $k$ to the number of clusters, $W(k)$ to the within cluster variation and $B(k)$ to the between cluster variation.

\begin{figure}
 \centering
 \includegraphics[width=\columnwidth]{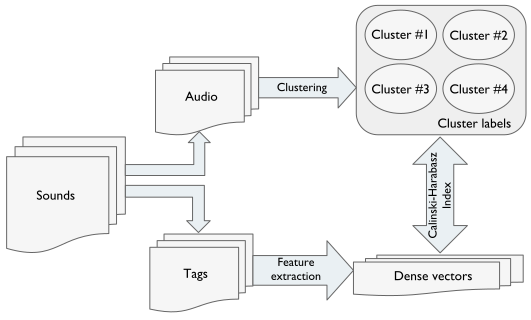}
 \caption{Diagram representing the steps of our internal evaluation making use of user-provided tags. The Calinski-Harabasz Index is calculated between the labels corresponding to the obtained clusters and the features derived from the sound tags. This evaluation is performed on the results of the 1000 most popular queries performed by Freesound users.}
 \label{fig:src_eval_tags}
\end{figure}

%
%
Instead of calculating this metric using the audio features used for clustering, we make use of the user-provided tags associated with the audio content as an external information.
This allows us to evaluate the overall quality of a clustering, from a semantic perspective.
From the tags associated to the content, we derive a feature using a Vector Space Model representation~\cite{salton1989automatic}.
This feature is a high-dimentional sparse vector where a value of 1 in one dimension refers to the presence of a specific tag.
We only consider the 5000 tags that occur the most in the overall Freesound collection.
We then reduce the size of this vector to 100, by applying Latent Semantic Analysis, which can capture synonymy relations~\cite{deerwester1990indexing}.
%
Due to the nature of tags, the validity metric we use is not always accurate.
In order to mitigate this problem, we average this metric on clusterings performed on the results of the 1000 most popular queries in Freesound.
In total, approximately 80k different sounds were used in this evaluation.
Figure \ref{fig:src_eval_tags} represents the evaluation pipeline.
%
The statistics are presented in Table \ref{tab:evaluation}.
%

\subsection{External Validation}
As an additional evaluation, we also make use of an
external validation metrics, which relies on known ground truth labels. 
We exploit data gathered within Freesound Annotator
~\cite{fonseca2017freesound} to construct 44 datasets comprising in total around 30k sounds and 215 different labels. 
Labels are drawn from the AudioSet Ontology~\cite{gemmeke2017audio}, which consists of a hierarchical taxonomy of 635 sound-related categories.
In our experiment, a dataset consists of one node in the taxonomy, and its labels are its direct children. 
This creates datasets of different sizes and with different levels of specificity. 
For instance, one broad dataset corresponds to natural sounds, containing the \textit{water}, \textit{wind}, \textit{thunderstorm} and \textit{fire} classes.
A more specific one contains only water sounds, with the \textit{rain}, \textit{stream}, \textit{steam}, \textit{waterfall}, \textit{gurgling}, and \textit{ocean} classes.
All the datasets contain sounds with a duration lower than 10 seconds and most of them contain only one salient source, which mitigates the inconvenience of using a statistical aggregation over the frame-based features.
Among the popular metrics used for comparing dataset partitions, the literature suggests that Adjusted Mutual Information (AMI) score~\cite{vinh2010information} is suited when the reference clustering (ground truth) is unbalanced and there exist small clusters~\cite{romano2016adjusting}.
This corresponds often to what we find in collaborative collections where the content is inconsistently distributed in terms of type and nature.
For evaluating the different sets of features with the different features, we perform clustering on the datasets and measure the similarity between the real partition (given by the ground truth labels) and the one given by the clustering methods by computing the Mutual Information score adjusted for chance (AMI), calculated as following:
\begin{equation}
AMI(U, V) = \frac{MI(U, V) - E\{MI(U, V)\}}{max\{H(U), H(V)\} - E\{MI(U, V)\}}
\end{equation}
\begin{equation}
MI(U, V) = \sum_{i} \sum_{j} P(i, j) \log(\frac{P(i, j)}{P(i) P(j)})
\end{equation}
Where $U$ and $V$ are the two partitions to compare, $H(U)$ and $H(V)$ their associated entropy. $P(i)$ and $P(j)$ the probabilities that a point belongs to cluster $U_i$ or $V_j$ respectively, $P(i,j)$ the probability that a point belongs to both cluster $U_i$ and $V_j$ from $U$ and $V$ respectively, and $E\{MI(U, V)\}$ corresponds to the expected mutual information between two random clusterings.
The Mutual Information metric (MI) quantifies the information shared by the two partitions and therefore can be used as a clustering similarity measure.
When adjusted for chance, the metric takes a value of 1 for two identical partitions and the value of 0 for two randomly dissimilar partitions.
%
Table \ref{tab:scores} shows some statistics of this score for the different audio features.

\subsection{Results}
In both evaluations, AudioSet embeddings lead to the best clustering performance.
This shows that novel deep learning approaches can produce semantically meaningful features that outperform traditional hand-crafted features for the unsupervised classification of sounds.
There is not any meaningful difference when applying manual feature selection over hand-crafted features (F1) motivated by results taken from the literature compared to using a large set of lowlevel features (F2).

Our approach for discarding low quality clusters using the confidence measure described in Section 3.3 shows little but consistent improvement with all the features in both experiments.

Another conclusion is that our proposed internal validation which makes use of accompanying tags provided by the users of the platform can give similar results as an external validation using ground-truth labels.
This provides a valuable framework for evaluating clustering algorithms in existing multimedia collections at scale, without needing to manually annotate a large amount of data.


\begin{table}
\small
 \begin{center}
 \begin{tabular}{l|cc|cc}
  \toprule
  \multirow{3}{*}{Features} & \multicolumn{4}{c}{CHI} \\
  & \multicolumn{2}{c}{no pruning} & \multicolumn{2}{c}{pruning} \\
  & mean & std & mean & std \\
  \midrule
  F1    & 3.36 & 5.87 & 3.88 & 7.25 \\
  F2    & 3.44 & 6.37 & 3.86 & 7.07 \\
  F3    & \textbf{4.29} & 6.82 & \textbf{5.29} & 11.06 \\
  \bottomrule
 \end{tabular}
\end{center}
 \caption{Clustering validity score (Calinski-Harabasz Index) using the different feature sets. Mean and standard deviation is calculated on the performance of the clustering of the results from the top 1000 most popular queries in Freesound. The pruning column corresponds to the validity score when discarding the cluster with the lowest confidence score defined in Section 3.3.}
 \label{tab:evaluation}
\end{table}

\begin{table}
\small
 \begin{center}
 \begin{tabular}{l|cc|cc}
  \toprule
  \multirow{3}{*}{Features} & \multicolumn{4}{c}{AMI} \\
  & \multicolumn{2}{c}{no pruning} & \multicolumn{2}{c}{pruning} \\
  & mean & std & mean & std \\
  \midrule
  F1    & 0.16 & 0.08 & 0.18 & 0.10 \\
  F2    & 0.15 & 0.09 & 0.16 & 0.12 \\
  F3    & \textbf{0.20} & 0.10 & \textbf{0.21} & 0.11 \\
  \bottomrule
 \end{tabular}
\end{center}
 \caption{Average performance (AMI) across the different dataset with the different features. An AMI close to 0 corresponds to a random partition while perfect matches gives 1 AMI. The pruning column corresponds to the performance when discarding the cluster with the lowest confidence score defined in Section 3.3.}
 \label{tab:scores}
\end{table}






\section{User evaluation}

In this section, we present our user-centered design process on the development of an interface for browsing sounds from large databases using the proposed clustering engine.
In this experiment, we use the AudioSet features (F3), which achieved the best performance in our comparative performance evaluation.
We use our interface prototypes as \textit{technology probes} to observe their use in a real context, to evaluate their functionalities and to inspire new ideas~\cite{hutchinson2003technology}. 

\subsection{Methodology}
We performed experiments with 4 users that are experienced with audio and sound design tasks, which is sufficient for detecting a large amount of usability problems~\cite{Nielsen_2000_5UserTest}.
All the participants were presented with the two searching tools (clustering facets filtering and 2D-visualisation) presented in Section 3.4.
The task consisted in gathering all the audio content needed in order to build the soundtrack of a short video, available at: \url{https://vimeo.com/333837958}. This video was chosen because it was short but presenting a lot of variety of elements to sonify.
The original sound was removed from the video.
Some guidelines were shown to the participants, together with verbal explanations given by the examinator who was present during the entire experiment.
At the end of the task, they were provided with a questionnaire containing some usability and engagement questions.
Finally, semi-structured interviews were carried out, including open-ended questions as well as specific questions related to observed behaviors during the performance of the task.
This enables discussion using thematic analysis in order to identify emerging themes from participants' answers.

\subsection{Results and discussion}

All the participants started by watching the video and noting down the concepts they would then look for within the audio collections.
Then, they started to use the search engine to look for the identified concepts needed for sonifying the video.
After entering a query, users often had a quick look at the top results.
They explained that it allowed them to figure out if the content retrieved was the one they were expecting. 
They then had the choice to either reformulate their query, or explore the retrieved results.
%
They found particularly useful the labels associated with each clusters in order to identify what kind of sounds were present in the results and what type of content each cluster contained.
%
They were either applying a cluster filtering to then browse the results in the retrieved sounds, or they would use terms from the cluster labels to reformulate their query.

However, some participants complained that the cluster labels were sometimes inappropriate, because they contained too broad concepts, or they were very similar for different clusters.
In these cases, it was hard to understand what type of content each cluster contained.
Nevertheless, the 2D-visualisation was then particularly helpful.
The participants were often listening to many sounds in a short amount of time thanks to the the fact that they only needed to hover the mouse on different nodes to start hearing some sounds. 
Whereas in the flat ranked list, they would have to manually trigger the players by clicking many times.
In the 2D-visualisation, their strategy was to first listen to a few dispersed sounds to quickly get an idea of how the sounds were organised in the space and what type of content was present in each cluster. 
Then, they would start exploring specific regions of interest, until they satisfy their need by retrieving one or several relevant sounds.
In addition, the users were often searching in a sound’s neighborhood. 
They explained that they wanted to find some slightly different variations of a relevant sound they already located.
However, understanding what the dimensions were capturing in the space was difficult.
The graph representation of the sound results was failing to reflect timbral characteristics of the sounds in a clear way.
Moreover, even if the graph was presenting a clear structure, it was not easy to understand to what it was corresponding to and to locate all the relevant content.
As a solution, a participant wanted to be able to select any retrieved sound from the ranked list and locate it in the 2D-visualisation.
This way, he explained that he could easily switch from one interface to the other, allowing him to efficiently combine the two interaction approaches.

The clustering engine was not very beneficial when the participants were using precise queries containing multiple words. 
They explained that in the context of sound design, they often know exactly what they need. And therefore they were often able to formulate a precise text query retrieving very specific content.
However, one drawback is that sounds that would not present the query terms as metadata would not be retrieved.
A solution to deal with bad recall performance of the system would be to use the audio-based representation to expand the retrieved results with sounds that are similar to the one retrieved.
In its current state, the prototype only applies clustering on the retrieved results, but does not include sounds that could have been relevant, but were missed to be retrieved by the text-based search engine.
Using the audio-based features for expanding the retrieved results would be interesting to study for queries that retrieved very few results. 

Some participants criticised the fact that the 2D-visualisation did not provide any representation of the waveform or any time-related information regarding the audio clips.
This made it hard to explore some results, as many of the users actively use waveforms in order to identify for instance if a foreground sound would appear at some time in the audio clip.
For that reason, in the 2D-visualisation, many participants were skipping some audio clips because the main acoustic event was not starting at the beginning of the clip.
Moreover, some participants said that they often use the waveform representation to assess some characteristics of the sounds, such as its dynamics, or the level of background noise present in the audio clip. 
Displaying the waveform with a time progression cursor of the current sound being played is therefore a key feature that would make the 2D-visualisation more useful and attractive.

Finally, in its current state, the clustering algorithm is able to discover distinctive and coherent groups in search results for many given queries.
However, in some cases, the quality of the clustering is still low, which made some users wasting time exploring bad clusters. 
Moreover, they explained that spending time exploring non-relevant clusterings could make them lose trust in the system, and therefore make them not use it again.
It was discussed in the interview the idea of reporting an estimate of the quality of the clustering, so that the user would be aware of its poor result and would be more confident while using the system. 
Using a confidence measure such as the one proposed in Section 3.3 could be a solution, or using directly the modularity of the graph partition.
Our evaluation of the strategy for discarding low quality clusters in Section 4 indicates that such measures can reflect, to some extent, the quality of the clustering.

\section{Conclusion}

In this paper we present a Search Result Clustering approach for enabling users to browse large online sound collections. To our knowledge, it is the first time that such an approach is applied for sound retrieval tasks.
%
We perform audio clustering using a graph-based approach.
%
We perform two evaluations for comparing the performance of different features.
The first one uses data of an online collection for accessing the performances at scale, whereas the second makes use of ground-truth labels from a reduced-size manually annotated collection.
We also propose a graph-based heuristic which allows to discard clusters of low quality.
For enabling the user to interact with the clusters, we proposed two approaches.
One consists on applying filters on the retrieved results, and the other involves a 2D-visualisation of a graph representation of the sounds.
We finally evaluate the system with users on a sound design task.

Results show that embeddings obtained by training neural networks on a supervised classification task with large amount of data can be used as a feature that increases the performance of the clustering compared to more traditional hand-crafted features.
Our experiment involving two evaluations indicates that using data from an existing online collection enables to evaluate clustering methods without the need to manually annotate content with pre-defined labels.
Testing the implemented prototype with users in a real-world scenario suggests that Search Result Clustering can assist the browsing of large sound collections.
Interesting feedback was gathered which will guide our future development of the clustering engine and its integration within the Freesound platform.
Furthermore, the methodology followed in this work provides a valuable framework for developing and evaluating clustering engines in the broad area of multimedia content retrieval.

%


\bibliographystyle{ACM-Reference-Format}
\bibliography{bib}

\end{document}